%% file: PAPS-AI.tex
\documentclass{llncs}
\usepackage{llncsdoc}

\usepackage{amsmath,amsfonts,amssymb,epsfig,epstopdf,url,array}

\usepackage{amsthm}
\usepackage{color,soul}
\usepackage{algorithmic}
\usepackage{balance}
\usepackage[ruled]{algorithm2e}
\usepackage{stmaryrd}
\usepackage[table]{xcolor}
\usepackage{centernot}
\usepackage{tikz}
\usepackage{booktabs}
\usepackage{pdfpages}
\usepackage{graphicx}
\usepackage{subfigure}
\usepackage[font=small,skip=8pt]{caption}
\usepackage{tabularx}
\usepackage{comment}
\usepackage{multirow}
\usepackage{multicol}
\usepackage{bigstrut}
\usepackage{pdflscape}
\usepackage{rotating}
\usepackage{float}
\usepackage{lipsum}
\usepackage{enumitem}
\usetikzlibrary{plotmarks,shapes,arrows,chains,hobby,backgrounds,calc,trees}
\usepackage{makecell}
\usepackage{dblfloatfix}    
\usepackage{pdfcomment} 

\definecolor{lightgray}{gray}{0.9}

\theoremstyle{definition}

\theoremstyle{definition}

\theoremstyle{definition}

\begin{document}

\mainmatter              
\title{PAPS: A Scalable Framework for Prioritization and Partial Selection of Security Requirements}
\titlerunning{PAPS}  
\author{Davoud Mougouei}
\authorrunning{Davoud Mougouei et al.} 

\institute{School of Computer Science, Engineering, and Mathematics\\ Flinders University, Adelaide, Australia}

\maketitle              

\begin{abstract}
\input{abstract}
\end{abstract}

\input{introduction}
\input{prepas}
\input{pas}

\input{pas_pre}
\input{pas_prior}
\input{pas_prior_fuzz}
\input{pas_prior_inference}
\input{pas_partial}

\input{conclusion}
\bibliographystyle{splncs}
\bibliography{ref}
\end{document}

%% file: abstract.tex
Owing to resource constraints, the existing prioritization and selection techniques for software security requirements (countermeasures) find a subset of higher-priority security requirements ignoring lower-priority requirements or postponing them to the future releases. Ignoring or postponing security requirements however, may on one hand leave some of the security threats (vulnerabilities) unattended and on the other hand influence other security requirements that rely on the ignored or postponed requirements. To address this, we have proposed considering partial satisfaction of security requirements when tolerated rather than ignoring those requirements or postponing them to the future. In doing so, we have contributed a goal-based framework that enables prioritization and partial selection of security requirements with respect to security goals. The proposed framework helps reduce the number of ignored (postponed) security requirements and consequently reduce the adverse impacts of ignoring security requirements in software products.

%% file: introduction.tex
\section{Introduction}
\label{introduction}
Security requirements (countermeasures) are to enhance security of software products. Nonetheless, due to the resources limitations it is hardly if ever possible to implement the entire set of identified security requirements for a software system \cite{loer2006integrated}. Consequently, an efficient prioritization and selection technique is required to find an optimal subset of security requirements for software products\cite{roy2012scalable,tonella2013interactive}. However, due to the following problems existing \textit{Prioritization And Selection} (PAS) techniques \cite{laurent2007towards,buyukozkan2005group,wiegers1999first,karlsson1997managing,karlsson1996software,karlsson1997cost,mohamed2008towards} have failed to be widely adopted by software practitioners \cite{ejnioui2012software,herrmann2008requirements}. The first problem i.e. \textit{Complexity} is that most of the existing PAS techniques are impractical to large number of requirements \cite{roy2012scalable,ejnioui2012software,berander2005requirements}. Adopted in prioritization of security requirements \cite{karlsson1998evaluation,mead2006identifying}, Analytic Hierarchy Process (AHP) has been the most promising \cite{karlsson1998evaluation} prioritization and selection technique \cite{ejnioui2012software} for years. However AHP suffers from high number of required comparisons \cite{karlsson1998evaluation}. Industrial studies have demonstrated that using AHP is not practical for more than $20$ requirements \cite{lehtola2004empirical}.There have been few reported techniques \cite{ejnioui2012software,berander2005requirements} to reduce the complexity of prioritization process but they have sacrificed the precision and consistency of the process \cite{ejnioui2012software}. 

The second problem is \textit{Imprecision} of PAS techniques \cite{ejnioui2012software} resulted by neglecting partiality of security and assuming that a security requirement can either be fully satisfied or ignored. But, ignoring (postponing) security requirements even if they are of lower priorities may leave some of security threats unattended and eventually result in security breaches in software systems. Moreover, ignoring a security requirement may also negatively influence the efficiency of other security requirements which rely on that requirement. As such, ignoring a security requirement may potentially cause cascading security vulnerabilities in software systems. 

To address the complexity and imprecision problems, we have proposed caring for partiality of security in a prioritization and selection process. In doing so, we have contributed a goal-based framework referred to as the \textit{Prioritization And Partial Selection} (PAPS) which enhances precision of a prioritization and selection process by allowing for partial selection (satisfaction)~\cite{mougouei2015partial} of security requirements when tolerated rather than ignoring those requirements altogether or postponing them to the future releases. The proposed framework helps reduce the number of ignored or postponed security requirements which will ultimately reduce the number of unattended security threats and mitigate the adverse impact of ignoring (postponing) security requirements. The PAPS framework is scalable to large number of requirements and allows for prioritization and selection of security requirements with respect to security goals. 

\begin{figure}[h!]
	\centering
	\centerline{\includegraphics[scale=0.7]{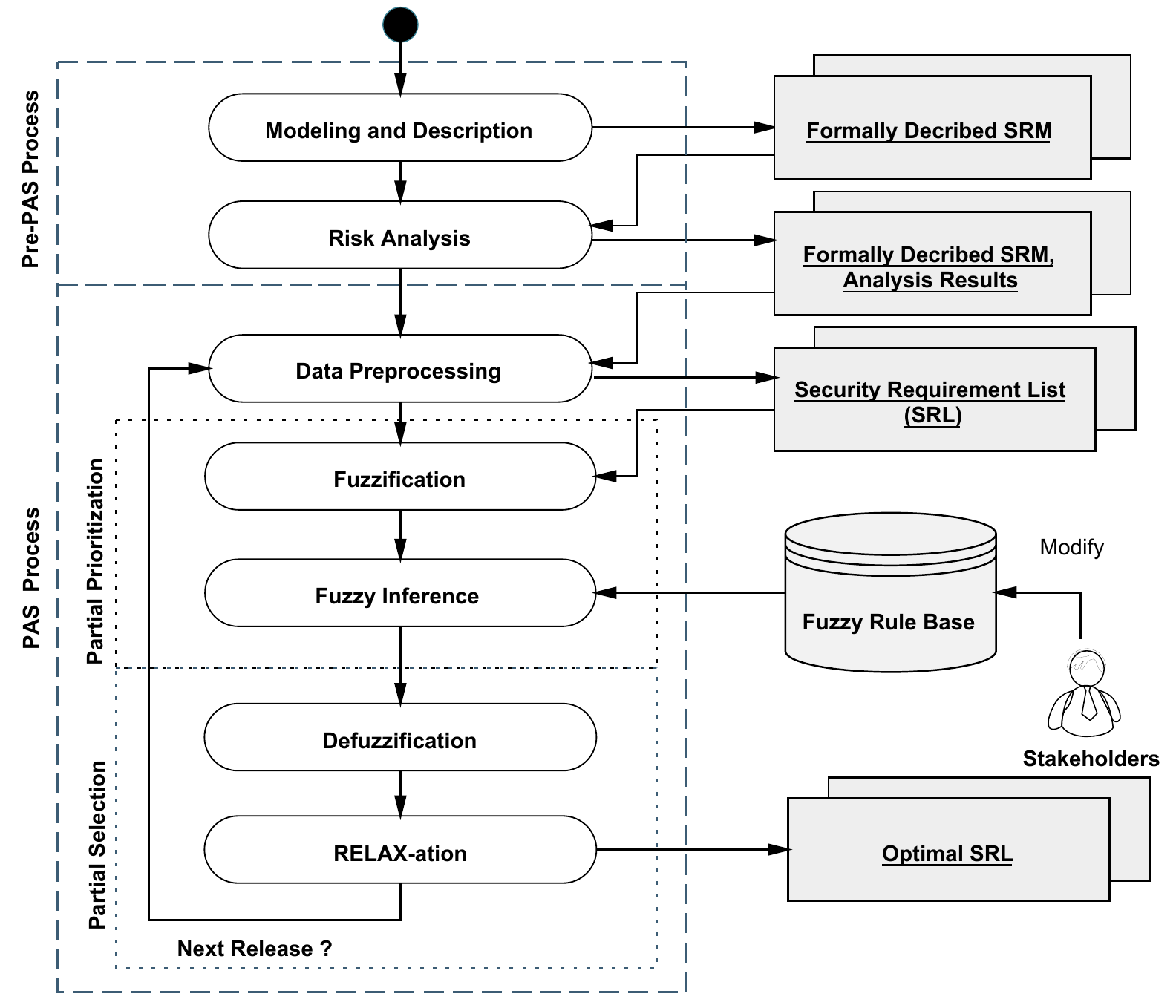}}
	\caption{Architecture of the Proposed PAPS Framework}
	\label{fig_paps}
\end{figure}

The proposed PAPS framework is composed of two major processes as demonstrated in Figure~\ref{fig_paps}. The first process referred to as \textit{Pre Prioritization and Selection} (Pre-PAS) includes modeling, description, and analysis of security requirements. The Pre-PAS process uses our previously developed security model known as \textit{Security Requirement Model} (SRM) \cite{mougouei2013goal} to capture partiality of security requirements (goals)~\cite{loer2006integrated}. The second process referred to as \textit{Prioritization and Selection} (PAS) process on the other hand, takes the formally described SRM of a software and its corresponding analysis results as the input and constructs the Security Requirement List (SRL) of that software. Then security requirements in SRL will be prioritized using a \textit{Fuzzy Inference System} (FIS) \cite{klir1995fuzzy}. Each security requirement in SRL contributes to satisfaction of at least one security goal. The FIS, infers linguistic priority values of the security requirements with respect to their impact, cost and technical-ability. We care for partiality in the optimal SRL through RELAX-ing \cite{whittle2010relax} satisfaction conditions of security requirements when partial satisfaction of those requirements is tolerated. Security requirements then will be RELAX-ed and included in the optimal SRL of software based on their fuzzy membership values.
 
The paper continues with an overview of our employed modeling and description technique (Section \ref{pas}). Then we introduce the PAPS framework (Section \ref{pas_pre}) and verify its validity through applying it to an Online Banking System (OBS)~\cite{mougouei2013goal}. The paper will be concluded in Section \ref{conclusion} with a summary of the work and some general remarks.  

%% file: prepas.tex
\section{Pre-PAS Process}
\label{prepas}
The Pre-PAS process as depicted in Figure~\ref{fig_paps}, includes modeling, description and analysis of security requirements. 

\subsection{Modeling and Description of Security Requirements}
\label{prepas_modeling_description}

\begin{figure}[!h]
	\centering
	\centerline{\includegraphics[scale=0.525]{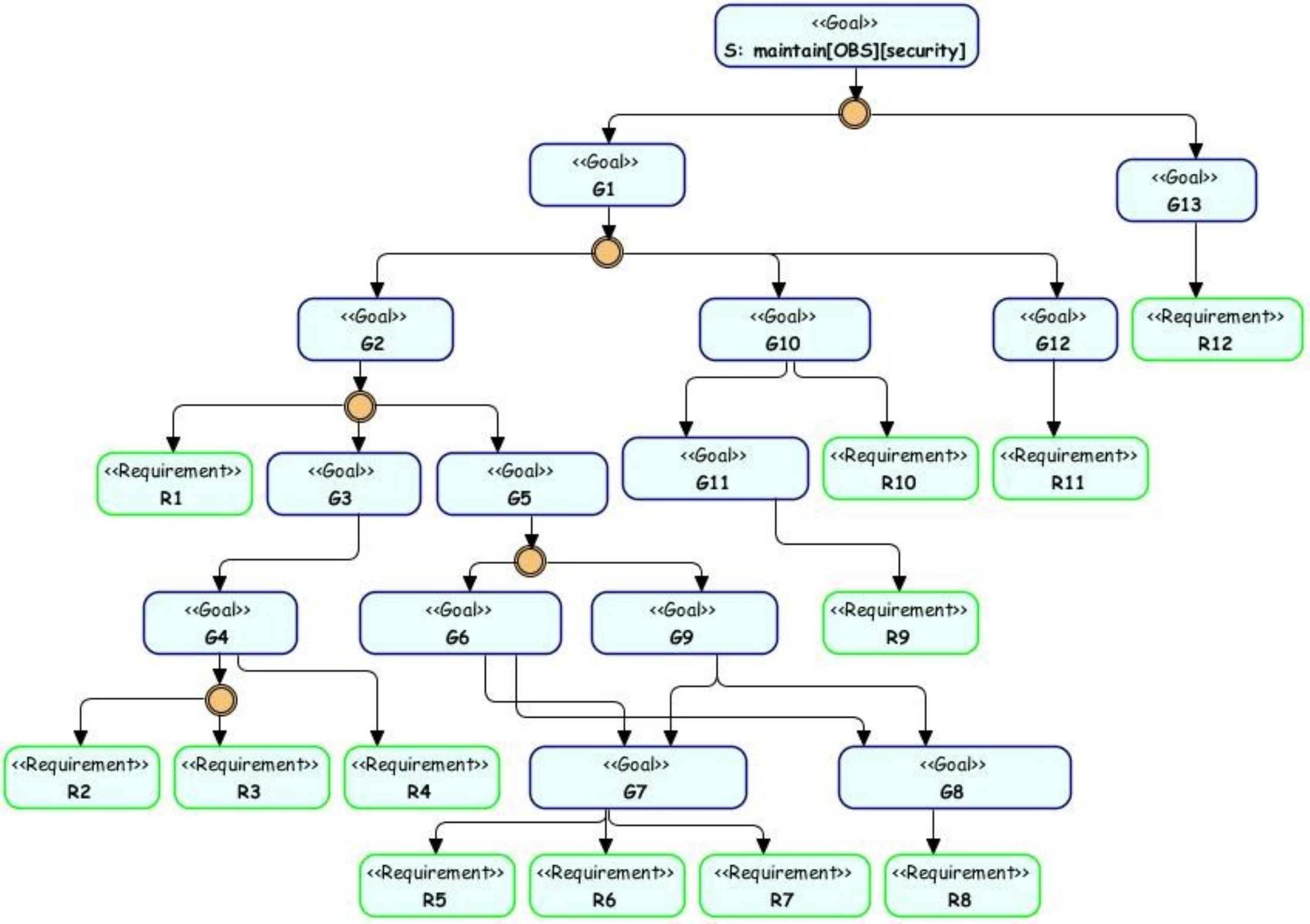}}
	\caption{SRM of the OBS}
	\label{fig_srm}
\end{figure}

Security Requirement Model (SRM) of a software is constructed by a goal-based modeling process presented in our earlier work\cite{mougouei2013goal}. The process starts with identification of the assets \cite{mead2006security,mougouei2013s} for a software product. Then, security goals will be developed to protect the assets against attack scenarios \cite{sindre2005eliciting}. Throughout the subsequent steps a security requirement model (SRM) of software will be constructed to mitigate security faults. Our goal-based modeling process made use of a combination of RELAX \cite{whittle2010relax} and KAOS \cite{van2004elaborating} description languages to describe security goals (requirements). The security requirement model of the OBS is illustrated in Figure \ref{fig_srm} and SRM nodes are described in Table \ref{table_description}. We have further, made use of a fuzzy-based technique presented in our earlier work \cite{mougouei2013fuzzyBased} to formally describe partiality in security requirement model of a software. In the following we have briefly described the main components of the employed fuzzy-based technique. 

\begin{table*}[t]
\caption{KAOS Description for Security Requirements (Goals) of the OBS.}
\label{table_description}
\centering
\input{table_description}
\end{table*}

\textit{i.Goal-Based Fuzzy Grammar (GFG)}. A GFG is defined as a quintuple of $GR = (G, R, P, S, \mu)$ in which $G$ is a set of security goals, $R$ is a set of security requirements, $P$ is a set of fuzzy derivation rules and $\mu$ denotes the membership function of derivation. $S$ represents the top-level security goal of the system. For $OBS$, $G=\{G_{1},...,G_{13}\}$, $R=\{R_{1},...,R_{12}\}$, $P=\{P_{1},...,P_{20}\}$ and \textit{S=``maintain [OBS] [security]''}. Due to its fuzziness, $GFG$ can properly capture partiality in SRM of a software. 

The elements of $P \in GR $ are expression of form given in \ref{Eq_rule} where `d' is the degree of contribution of a sub-goal `w' to satisfaction of a goal `r'. If $r_{1},..., r_{n}$ are fuzzy statements in $(G \cup R)^*$ and $r_{1} \rightarrow r_{2} \rightarrow ... \rightarrow r_{n}$, then we call this chain as goal derivation chain under the employed $GFG$. 

\begin{align}
\label{Eq_rule}
& \mu(r\rightarrow w)=d , d\in[0,1] \text{ or } \mu(r,w)=d
\end{align}

\begin{table*}[t]
	\caption{Derivation rules for SRM of the OBS.}
	\label{table_rules}
	\centering
	\input{table_rules}
\end{table*}

\textit{ii. Extracting Derivation Rules}. The employed description technique, constructs a GFG for a given SRM and identifies the derivation rules~\cite{mougouei2013fuzzy,mougouei2013fuzzyBased,mougouei2013fuzzy1}. The degree to which the successive of a rule contributes to satisfaction of its predecessor, will specify its membership value. This value will be determined by the membership function µ of GFG \cite{mougouei2013fuzzyBased}. The extracted derivation rules for SRM of the OBS, and their corresponding membership values are listed in Table \ref{table_rules}. 


\subsection{Risk Analysis}
\label{prepas_risk} 

During the risk analysis, the cost and technical-ability of security requirements will be identified. 

\textit{Cost of Implementation}. Owing to budget limitations we need to care for cost of implementation while selecting and prioritizing security requirements \cite{karlsson1997cost}. Table \ref{table_risk} has listed the cost and values for security requirements in the SRM of the OBS. Cost of implementation is a real number in $[1,100]$ \cite{mougouei2012measuring,mougouei2012evaluating}.  

\begin{table*}[t]
	\caption{Cost and Technical-ability of the OBS Security Requirements.}
	\label{table_risk}
	\centering
	\input{table_risk}
\end{table*}

\textit{Technical-Ability}. Technical-ability is a real number in $[0,1]$ which reflects the ease of implementation for each requirement. Technical-ability of security requirements of the OBS are computed based on (\ref{Eq_technical}) and listed in Table \ref{table_risk}.  

\begin{align}
\label{Eq_technical}
& Technical\text{-}Ability=\frac{1}{\textit{Technical Complexity of Requirement}}
\end{align}

%% file: table_description.tex
\normalsize\resizebox {0.95\textwidth }{!}{
\begin{tabular}{llll}
	\toprule[1.5pt]
	\textbf{Goal} &
	\textbf{Description} &
	\textbf{Requirement} &
	\textbf{Description}
	\bigstrut\\
	\midrule[1.5pt]
	\cellcolor{blue!10}\cellcolor{blue!10}\textit{$S$} &
	\cellcolor{blue!10}\textit{maintain OBS security} &
	\multicolumn{1}{l}{\cellcolor{green!10}$R_{1}$} &
	\cellcolor{green!10}\textit{achieve request transaction code}
	\bigstrut[t]\\
	\cellcolor{blue!10}$G_{1}$ &
	\cellcolor{blue!10}\textit{avoid  transfer money out of account} &
	\multicolumn{1}{l}{\cellcolor{green!10}$R_{2}$} &
	\cellcolor{green!10}\textit{achieve latency examination }
	\\
	\cellcolor{blue!10}$G_{2}$ &
	\cellcolor{blue!10}\textit{avoid unauthorized online transfer} &
	\multicolumn{1}{l}{\cellcolor{green!10}$R_{3}$} &
	\cellcolor{green!10}\textit{achieve one-time pad}
	\\
	\cellcolor{blue!10}$G_{3}$ &
	\cellcolor{blue!10}\textit{avoid stealing id and password} &
	\multicolumn{1}{l}{\cellcolor{green!10}$R_{4}$} &
	\cellcolor{green!10}\textit{achieve SSL}
	\\
	\cellcolor{blue!10}$G_{4}$ &
	\cellcolor{blue!10}\textit{avoid man in the middle} &
	\multicolumn{1}{l}{\cellcolor{green!10}$R_{5}$} &
	\cellcolor{green!10}\textit{achieve password trial limitation}
	\\
	\cellcolor{blue!10}$G_{5}$ &
	\cellcolor{blue!10}\textit{avoid guessing id and password} &
	\multicolumn{1}{l}{\cellcolor{green!10}$R_{6}$} &
	\cellcolor{green!10}\textit{achieve password policy}
	\\
	\cellcolor{blue!10}$G_{6}$ &
	\cellcolor{blue!10}\textit{avoid dictionary attack} &
	\multicolumn{1}{l}{\cellcolor{green!10}$R_{7}$} &
	\cellcolor{green!10}\textit{achieve password encryption}
	\\
	\cellcolor{blue!10}$G_{7}$ &
	\cellcolor{blue!10}\textit{avoid guess password} &
	\multicolumn{1}{l}{\cellcolor{green!10}$R_{8}$} &
	\cellcolor{green!10}\textit{achieve random id}
	\\
	\cellcolor{blue!10}$G_{8}$ &
	\cellcolor{blue!10}\textit{avoid guess id } &
	\multicolumn{1}{l}{\cellcolor{green!10}$R_{9}$} &
	\cellcolor{green!10}\textit{achieve CAPTCHA}
	\\
	\cellcolor{blue!10}$G_{9}$ &
	\cellcolor{blue!10}\textit{avoid brute forcing} &
	\multicolumn{1}{l}{\cellcolor{green!10}$R_{10}$} &
	\cellcolor{green!10}\textit{achieve complex pin}
	\\
	\cellcolor{blue!10}$G_{10}$ &
	\cellcolor{blue!10}\textit{avoid unauthorized transfer via debit card} &
	\multicolumn{1}{l}{\cellcolor{green!10}$R_{11}$} &
	\cellcolor{green!10}\textit{achieve access control}
	\\
	\cellcolor{blue!10}$G_{11}$ &
	\cellcolor{blue!10}\textit{maintain transfer network security} &
	\multicolumn{1}{l}{\cellcolor{green!10}$R_{12}$} &
	\cellcolor{green!10}\textit{achieve redundant server}
	\\
	\cellcolor{blue!10}$G_{12}$ &
	\cellcolor{blue!10}\textit{avoid hijack server} &
	\multicolumn{1}{l}{\cellcolor{green!10}\textit{}} &
	\cellcolor{green!10}\textit{}
	\\
	\cellcolor{blue!10}$G_{13}$ &
	\cellcolor{blue!10}\textit{maintain service availability} &
	\multicolumn{1}{l}{\cellcolor{green!10}\textit{}} &
	\cellcolor{green!10}\textit{}
	\bigstrut[b]\\
	\bottomrule[1.5pt]
\end{tabular}}%

%% file: table_rules.tex
\normalsize\rowcolors{1}{}{lightgray}
\resizebox {0.75\textwidth }{!}{
\begin{tabular}{llll}
	\toprule[1.5pt]
	\textbf{Rule}  &
	\textbf{Membership Value\quad} &
	\textbf{Rule} &
	\textbf{Membership Value}
	\bigstrut\\
	\midrule
	\textit{$P_{1}:\phantom{s} S \rightarrow G{1}G_{13}$ } &
	\phantom{ss}\textit{$ 0.95 $} &
	\textit{$P_{11}:\phantom{s} G_{4} \rightarrow R_{2}R_{3}$} &
	\phantom{ss}\textit{$ 0.75 $}
	\bigstrut[t]\\
	\textit{$P_{2}:\phantom{s} G_{1} \rightarrow G_{2}G_{10}G_{12}$} &
	\phantom{ss}\textit{$ 0.95 $} &
	\textit{$P_{12}:\phantom{s} G_{4} \rightarrow R_{4}$} &
	\phantom{ss}\textit{$ 0.9 $}
	\\
	\textit{$P_{3}:\phantom{s} G_{13} \rightarrow R_{12}$} &
	\phantom{ss}\textit{$ 0.9 $} &
	\textit{$P_{13}:\phantom{s} G_{6} \rightarrow G_{7}$} &
	\phantom{ss}\textit{$ 0.6 $}
	\\
	\textit{$P_{4}:\phantom{s} G_{2} \rightarrow R_{1}G_{3}G_{5}$} &
	\phantom{ss}\textit{$ 0.85 $} &
	\textit{$P_{14}:\phantom{s} G_{6} \rightarrow G_{8}$} &
	\phantom{ss}\textit{$ 0.6 $}
	\\
	\textit{$P_{5}:\phantom{s} G_{10} \rightarrow G_{11}$} &
	\phantom{ss}\textit{$ 0.9 $} &
	\textit{$P_{15}:\phantom{s} G{9} \rightarrow G_{7}$} &
	\phantom{ss}\textit{$ 0.65 $}
	\\
	\textit{$P_{6}:\phantom{s} G_{10} \rightarrow R_{10}$} &
	\phantom{ss}\textit{$ 0.4 $} &
	\textit{$P_{16}:\phantom{s} G_{9} \rightarrow G_{8}$} &
	\phantom{ss}\textit{$ 0.6 $}
	\\
	\textit{$P_{7}:\phantom{s} G_{12} \rightarrow R_{11}$} &
	\phantom{ss}\textit{$ 0.9 $} &
	\textit{$P_{17}:\phantom{s} G_{7} \rightarrow R_{5}$} &
	\phantom{ss}\textit{$ 0.7 $}
	\\
	\textit{$P_{8}:\phantom{s} G_{3} \rightarrow G_{4}$} &
	\phantom{ss}\textit{$ 0.85 $} &
	\textit{$P_{18}:\phantom{s} G_{7} \rightarrow R_{6}$} &
	\phantom{ss}\textit{$ 0.8 $}
	\\
	\textit{$P_{9}:\phantom{s} G_{5} \rightarrow G_{6}G_{9}$} &
	\phantom{ss}\textit{$ 0.9 $} &
	\textit{$P_{19}:\phantom{s} G_{7} \rightarrow R_{7}$} &
	\phantom{ss}\textit{$ 0.9 $}
	\\
	\textit{$P_{10}:\phantom{s} G_{11} \rightarrow R_{9}$} &
	\phantom{ss}\textit{$ 0.8 $} &
	\textit{$P_{20}:\phantom{s} G_{8} \rightarrow R_{8}$} &
	\phantom{ss}\textit{$ 0.6 $}
	\bigstrut[b]\\
	\bottomrule[1.5pt]
\end{tabular}}%

%% file: table_risk.tex
\normalsize\rowcolors{1}{}{lightgray}
\resizebox {0.75\textwidth }{!}{
\begin{tabular}{lllllllllllll}
	\toprule[1.5pt]
	\textbf{Requirement} &
	\textit{$R_1$} &
	\textit{$R_2$} &
	\textit{$R_3$} &
	\textit{$R_4$} &
	\textit{$R_5$} &
	\textit{$R_6$} &
	\textit{$R_7$} &
	\textit{$R_8$} &
	\textit{$R_9$} &
	\textit{$R_{10}$} &
	\textit{$R_{11}$} &
	\textit{$R_{12}$}
	\bigstrut\\
	\hline
	\textbf{Cost} &
	\textit{$ 0.50 $} &
	\textit{$0.70$} &
	\textit{$0.70$} &
	\textit{$ 0.30 $} &
	\textit{$ 0.05 $} &
	\textit{$ 0.50 $} &
	\textit{$ 0.20 $} &
	\textit{$ 0.01 $} &
	\textit{$ 0.60 $} &
	\textit{$ 0.10 $} &
	\textit{$0.70$} &
	\textit{$ 1.00 $}
	\bigstrut[t]\\
	\textbf{Technical-ability} &
	\textit{$ 1.00 $} &
	\textit{$ 0.20 $} &
	\textit{$ 0.10 $} &
	\textit{$ 0.90 $} &
	\textit{$ 1.00 $} &
	\textit{$ 0.30 $} &
	\textit{$ 0.20 $} &
	\textit{$ 1.00 $} &
	\textit{$ 0.10 $} &
	\textit{$ 1.00 $} &
	\textit{$ 0.20 $} &
	\textit{$ 0.20 $}
	\bigstrut[b]\\
	\bottomrule[1.5pt]
\end{tabular}}%

%% file: pas.tex
\section{Prioritization and Selection Process}
\label{pas}

The PAS process starts with preprocessing FIS inputs. Preprocessing includes construction of SRL and calculation of impacts for security requirements in the SRL. Subsequently, impacts, costs and technical-abilities will be fuzzified \cite{bede2013fuzzy} to serve as the inputs of the FIS. A Mamdani-type \cite{mamdani1974application} fuzzy inference system then specifies the priorities of security requirements. Prioritization can be performed with special focus on satisfaction of a security goal. Finally, prioritized requirements will be partially selected (when tolerated) through RELAX-ation of their satisfaction conditions. To perform RELAX-ation, we need to obtain Required Degree of Satisfaction (RDS) for each security requirement through deffuzification of priority values. 

%% file: pas_pre.tex
\subsection{Data Preprocessing}
\label{pas_pre}


\setlength{\fboxsep}{6pt}%
\setlength{\fboxrule}{0.5pt}%

Data preprocessing includes construction of SRL and calculation of impact values. Let $GR = (G, R, P, S, \mu)$ be the GFG of a SRM. For each security goal $g \in G$, $SRL[g]$ contains security requirements which contribute to satisfaction of goal g. $SRL[g]$ is constructed for every goal \textit{g} in SRM of the system. This allows for goal-oriented prioritization of security requirements with special focus on satisfaction of individual security goals. For each security requirement $x$ in $SRL[g]$ the degree to which $x$ contributes to satisfaction of the goal $g$ is referred to as the impact of $x$ on $g$ and computed by taking maximum ($ \oplus $) over membership degree of all derivations paths that can derive $x$ from $g$. Membership of each path is computed by taking minimum ($ \otimes $) over membership degrees of all derivations rules on the path. Impacts for security requirements of the OBS are listed in Table \ref{table_impact}. 


\begin{align}
\label{Eq_imp}
& DC_{g} (x) = \mu_{g}(x)= \oplus(\mu(g,r_1)\otimes\mu(r_1,r_2)\otimes ... \otimes\mu(r_n,x))
\end{align}


\begin{table*}[t]
	\caption{Impact of Security Requirements in the SRM of the OBS.}
	\label{table_impact}
	\centering
	\input{table_impact}
\end{table*}

%% file: table_impact.tex
\rowcolors{1}{}{lightgray}
\resizebox {0.75\textwidth }{!}{\begin{tabular}{lllllllllllll}
\Xhline{4\arrayrulewidth}
Goal &
  $\mu(R_1)$ &
  $\mu(R_2)$ &
  $\mu(R_3)$ &
  $\mu(R_4)$ &
  $\mu(R_5)$ &
  $\mu(R_6)$ &
  $\mu(R_7)$ &
  $\mu(R_8)$ &
  $\mu(R_9)$ &
 $ \mu(R_{10})$ &
  $\mu(R_{11})$ &
  $\mu(R_{12}$)
  \\
\Xhline{4\arrayrulewidth}
S &
  $0.85$ &
  $0.75$ &
  $0.75$ &
  $0.85$ &
  $0.65$ &
  $0.65$ &
  $0.65$ &
  $0.60$ &
  $0.80$ &
  $0.40$ &
  $0.90$ &
  $0.90$
  \bigstrut\\
$G_1$ &
  $0.85$ &
  $0.75$ &
  $0.75$ &
  $0.85$ &
  $0.65$ &
  $0.65$ &
  $0.65$ &
  $0.60$ &
  $0.80$ &
  $0.40$ &
  $0.90$ &
  $0.00$
  \\
$G_2$ &
  $0.85$ &
  $0.75$ &
  $0.75$ &
  $0.85$ &
  $0.65$ &
  $0.65$ &
  $0.65$ &
  $0.60$ &
  $0.00$ &
  $0.00$ &
  $0.00$ &
  $0.00$
  \\
$G_3$ &
  $0.00$ &
  $0.75$ &
  $0.75$ &
  $0.85$ &
  $0.00$ &
  $0.00$ &
  $0.00$ &
  $0.00$ &
  $0.00$ &
  $0.00$ &
  $0.00$ &
  $0.00$
  \\
$G_4$ &
  $0.00$ &
  $0.75$ &
  $0.75$ &
  $0.90$ &
  $0.00$ &
  $0.00$ &
  $0.00$ &
  $0.00$ &
  $0.00$ &
  $0.00$ &
  $0.00$ &
  $0.00$
  \\
$G_5$ &
  $0.00$ &
  $0.00$ &
  $0.00$ &
  $0.00$ &
  $0.65$ &
  $0.65$ &
  $0.65$ &
  $0.60$ &
  $0.00$ &
  $0.00$ &
  $0.00$ &
  $0.00$
  \\
$G_6$ &
  $0.00$ &
  $0.00$ &
  $0.00$ &
  $0.00$ &
  $0.60$ &
  $0.60$ &
  $0.60$ &
  $0.60$ &
  $0.00$ &
  $0.00$ &
  $0.00$ &
  $0.00$
  \\
$G_7$ &
  $0.00$ &
  $0.00$ &
  $0.00$ &
  $0.00$ &
  $0.70$ &
  $0.80$ &
  $0.90$ &
  $0.00$ &
  $0.00$ &
  $0.00$ &
  $0.00$ &
  $0.00$
  \\
  $G_8$ &
  $0.00$ &
  $0.00$ &
  $0.00$ &
  $0.00$ &
  $0.00$ &
  $0.00$ &
  $0.00$ &
  $0.60$ &
  $0.00$ &
  $0.00$ &
  $0.00$ &
  $0.00$
  \\
$G_9$ &
  $0.00$ &
  $0.00$ &
  $0.00$ &
  $0.00$ &
  $0.65$ &
  $0.65$ &
  $0.65$ &
  $0.65$ &
  $0.00$ &
  $0.00$ &
  $0.00$ &
  $0.00$
  \\
$G_{10}$ &
  $0.00$ &
  $0.00$ &
  $0.00$ &
  $0.00$ &
  $0.00$ &
  $0.00$ &
  $0.00$ &
  $0.00$ &
  $0.80$ &
  $0.40$ &
  $0.00$ &
  $0.00$
  \\
$G_{11}$ &
  $0.00$ &
  $0.00$ &
  $0.00$ &
  $0.00$ &
  $0.00$ &
  $0.00$ &
  $0.00$ &
  $0.00$ &
  $0.80$ &
  $0.00$ &
  $0.00$ &
  $0.00$
  \\
$G_{12}$ &
  $0.00$ &
  $0.00$ &
  $0.00$ &
  $0.00$ &
  $0.00$ &
  $0.00$ &
  $0.00$ &
  $0.00$ &
  $0.00$ &
  $0.00$ &
  $0.90$ &
  $0.00$
  \\
$G_{13}$ &
  $0.00$ &
  $0.00$ &
  $0.00$ &
  $0.00$ &
  $0.00$ &
  $0.00$ &
  $0.00$ &
  $0.00$ &
  $0.00$ &
  $0.00$ &
  $0.00$ &
  $0.90$
  \\
\Xhline{4\arrayrulewidth}
\end{tabular}}%

%% file: pas_prior.tex
\subsection{Prioritization}
\label{pas_prior}

Prioritization starts with fuzzification of FIS inputs (impact, cost, and technical-ability of security requirements). As depicted in Figure \ref{fig_paps}, the fuzzified values will serve as the inputs for the FIS. FIS then infers the fuzziﬁed priority values of requirements based on fuzzy-rule-base of the PAPS framework. Priority of each requirement specifies the required satisfaction level of that requirement. 



%% file: pas_prior_fuzz.tex
\subsection{Fuzzification}
\label{pas_prior_fuzz}

FIS inputs (impact, cost and technical-ability) are categorized under three fuzzy categories: Low (L), Medium (M) and High (H). Consequently, three membership functions are defined for each input and its corresponding categories. We use a semi-trapezoids shape for membership functions (Figure \ref{fig_inference}). Hence, four diverse points are required to define each membership function as given by (\ref{Eq_mem1})-(\ref{Eq_mem3}). We have use a combination of Fuzzy Control Language (FCL) and jFuzzyLogic \cite{lewis1998programming} to implement the membership functions.


\begin{align}
	\label{Eq_mem1}
	& \mu_{iv}(x) = \max (\min(\frac{x-x_0}{x_1-x_0},1,\frac{x_3-x}{x_3-x_2}),0)\\
    \label{Eq_mem2}
   	& \mu(x_0)=\mu(x_3)=0,\mu(x_1)=\mu(x_2)=1 \\ 
	\label{Eq_mem3}
	& i \in \{impact,cost,technical\text{-}ability\} ,v \in \{low,medium,high\}
\end{align}



\begin{figure}[h!]
	\centering
	\centerline{\fbox{\includegraphics[scale=0.5]{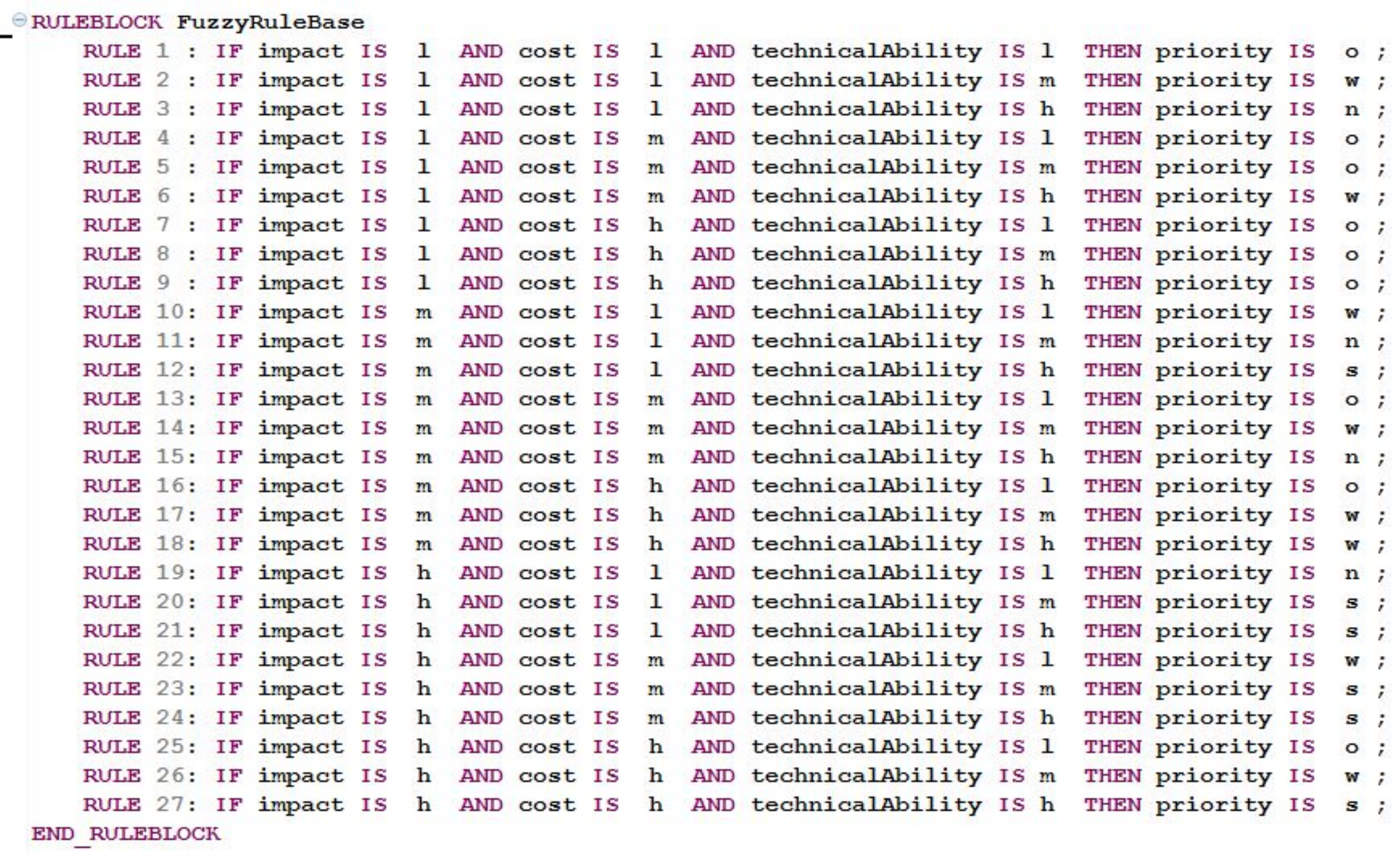}}}
	\caption{Fuzzy Rules Implemented in FCL}
	\label{fig_rulebase}
\end{figure}

%% file: pas_prior_inference.tex
\subsection{Fuzzy Inference}
\label{pas_prior_inference}

Priorities of security requirements are inferred by a Mamdani-Type \cite{mamdani1974application} FIS using the fuzzy rule-base of Figure~\ref{fig_rulebase}. In this regard, linguistic priorities Optional (O), Weak (W), Normal (N), or Strong (S) will be assigned to security requirements. Fuzzy rules of course can be tailored to the organizational and technical concerns of stakeholders. Priorities of the OBS security requirements are listed in Table~\ref{table_rds}. Each security requirement in Table~\ref{table_rds} is assigned $ 14 $ priority values each of which computed with regard to a specific security goal to assist goal-based prioritization and selection of requirements. Goal-based PAS provides structured arguments to the stakeholders. For instance, requirement ``$R_7$" in SRL of the OBS is strongly (weakly) required for satisfaction of the goal ``$G_2$" (``$G_6$"). 

\begin{figure}[!htbp]
	\centering
	\centerline{\includegraphics[scale=0.55]{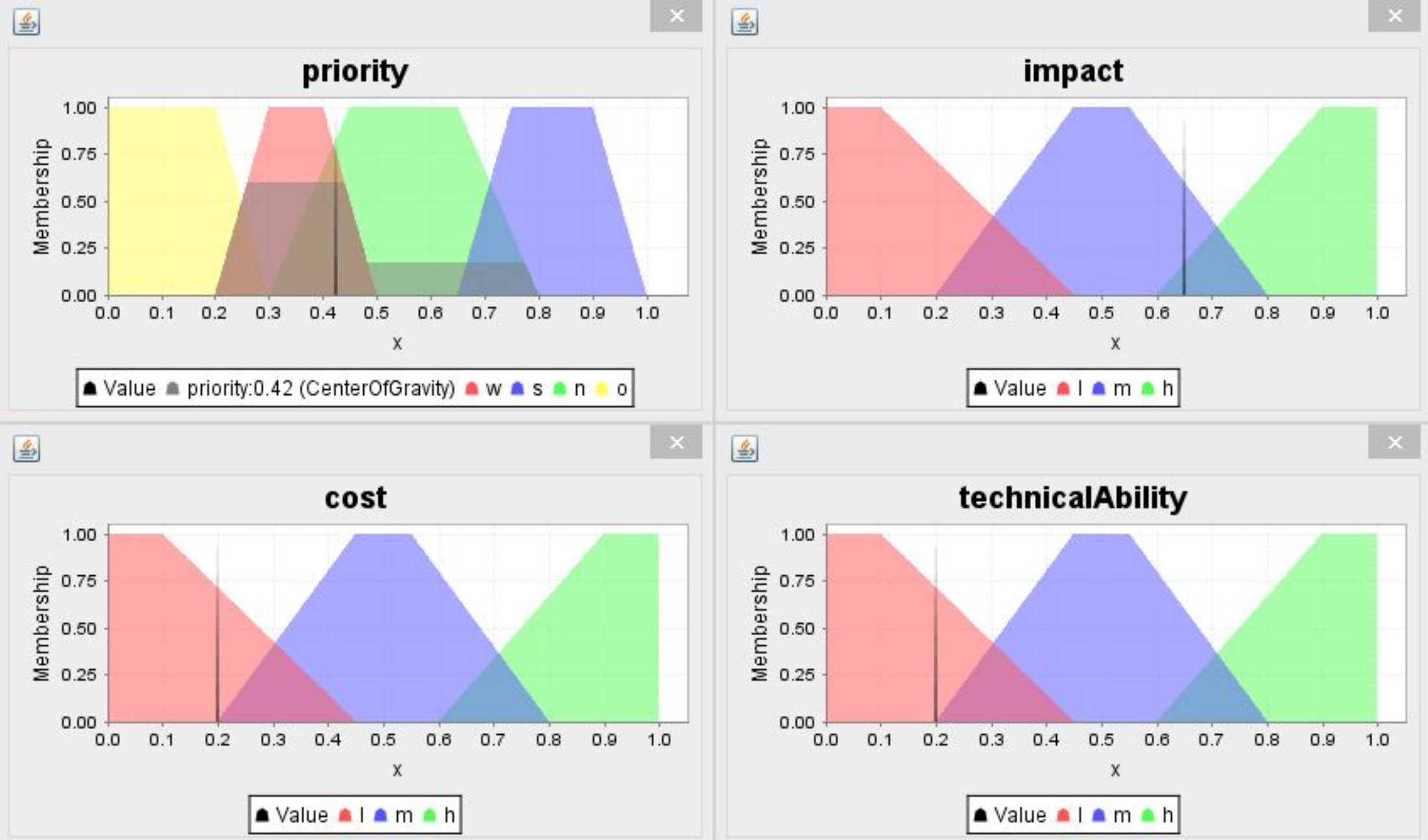}}
	\caption{Fuzzy Inference for $R_7$ with respect to top-level security goal S}
	\label{fig_inference}
\end{figure}
\vspace{-1em}

%% file: pas_partial.tex
\subsection{Partial Selection}
\label{pas_partial}

It is happening quite often in software systems that a security requirement cannot be either fully implemented or ignored. In such cases, security requirements may be tolerated \cite{whittle2010relax,cheng2009goal} to be partially satisfied (selected). Partial satisfaction of a security requirement can be explicitly addressed through RELAX-ing \cite{cheng2009goal} its satisfaction condition (level of satisfaction). For instance, implementation of a complex password policy in a software system increases the level of security on one hand and reduces the usability of the system \cite{adams1999users} on the other hand. Alternatively, implementing a less complex password policy may be tolerated to maintain usability of the system. In this case, the requirement will be partially selected through RELAX-ing its satisfaction condition. 

\begin{table*}[t]
	\caption{RDS (Linguistic Priority) of the Security Requirements of the OBS.}
	\label{table_rds}
	\centering
	\input{table_rds}
\end{table*}

\subsubsection{Defuzzification}
\label{partial_def}

Our employed RELAX-ation technique\cite{mougouei2013goal,mougouei2015partial} needs crisp values to specify the required level of satisfaction for RELAX-ed security requirements. Hence, defuzzification is required to map the linguistic priority values of requirements into their corresponding crisp values. To do so, we use the Center of Gravity (COG) \cite{van2006fast} formula given by (\ref{Eq_rds}) and the membership function of the priority (Figure~\ref{fig_inference}) to deffuzify the priority values. $ x $ in this equation denotes crisp priority values on the $x$ axis of the priority graph in Figure~\ref{fig_inference}. Also $\mu_{x}$ denotes the membership function of priorities. 

\begin{align}
\label{Eq_rds}
& \text{Defuzzified Priority}=\frac{\int_{0}^{1}\mu_{x}\times x\times d_{x}}{\int_{0}^{1}\mu_{x}\times d_{x}}
\end{align}
\vspace{-3em}
\begin{table}[!htbp]
	\caption{RELAX-ed Security Requirements of OBS.}
	\label{table_relaxed}
	\centering
	\input{table_relaxed}
\end{table}
\vspace{-2em}
\subsubsection{RELAX-ation}
\label{partial_relax}

A requirement is RELAX-ed by relaxing its satisfaction condition (level). For this purpose we use a RELAX-ation technique proposed in~\cite{whittle2010relax}. Fuzzy semantic of the RELAX statements in the technique used properly captures partiality of security requirements \cite{mougouei2013goal}. Equations (\ref{eq_relax1})-(\ref{eq_relax3}) demonstrate this logic. 

In this equation, the aim is to maximize for each requirement $R_i$ the value of the membership function $\mu(V_i - (RDS_i \times OV_i))$ which is equivalent to finding a value $V_i$ for the satisfaction condition of $R_i$ as close as possible to the relaxed value $RDS_i\times OV_i$ where $RDS_i$ denotes the required degree of satisfaction for $R_i$ and $OV_i$ is the optimal value of the satisfaction condition in the absence of all resource limitations. 



\begin{align}
\label{eq_relax1}
Maximize \phantom{s} &\mu \big(V_i - (RDS_i \times OV_i) \big),& i=1,...,n\\
\label{eq_relax2}
Subject\phantom{s}to\phantom{s} &\mu(x_j)\in [0,1], &j=1,...,n \\ 
\label{eq_relax3}
& \mu(0)= 1\\ \nonumber
\end{align}

For instance consider RELAX-ing from security requirements of the OBS ``$R_6$: achieve password policy" where based on Table~\ref{table_rds}, $ SRL[S][R6] $= `weak’. When a less complex password encryption can be tolerated in $ SRL[S] $, satisfaction condition of $R_6$ can be RELAX-ed as : ``$ R_6 $: System shall achieve password policy [complexity] as close as possible to ($ RDS_6 \times OV_6 $). The membership function $\mu$ is maximizes ($\mu(x)=1$) when the variance of the password complexity is equal to $RDS_6 \times OV_6$. In a similar way, security requirements in $ SRL[S] $ of the OBS are RELAX-ed and listed in Table~\ref{table_relaxed}. It is important to note that ‘relaxed’ attributes are constructed based on the satisfaction metrics of security requirements. As depicted in Table~\ref{table_relaxed}, satisfaction metrics are placed in the brackets. For instance, satisfaction metric of $ R_6 $ is ``complexity" of the password policy while ``length" of encryption key is measures satisfaction of $ R_7 $.

%% file: table_rds.tex
\rowcolors{1}{}{lightgray}
\resizebox {0.9\textwidth }{!}{
\begin{tabular}{lllllllllllll}
\Xhline{4\arrayrulewidth}
\multirow{2}[4]{*}{Goal} &
  \multicolumn{12}{c}{RDS Values}
  \bigstrut\\
\cline{2-13} &
  $R_1$ &
  $R_2$ &
  $R_3$ &
  $R_4$ &
  $R_5$ &
  $R_6$ &
  $R_7$ &
  $R_8$ &
  $R_9$ &
  $R_{10}$ &
  $R_{11}$ &
  $R_{12}$
  \bigstrut\\
\Xhline{4\arrayrulewidth}
S &
  0.82 (S) &
  0.25 (W) &
  0.25 (W) &
  0.82 (S) &
  0.59 (N) &
  0.36 (W) &
  0.42 (N) &
  0.55 (N) &
  0.35 (W) &
  0.55 (N) &
  0.25 (W) &
  0.13 (O)
  \bigstrut\\
$G_1$ &
  0.82 (S) &
  0.25 (W) &
  0.25 (W) &
  0.82 (S) &
  0.59 (N) &
  0.36 (W) &
  0.42 (N) &
  0.55 (N) &
  0.35 (W) &
  0.55 (N) &
  0.25 (W) &
  -
  \\
$G_2$ &
  0.82 (S) &
  0.25 (W) &
  0.25 (W) &
  0.82 (S) &
  0.59 (N) &
  0.36 (W) &
  0.42 (N) &
  0.55 (N) &
  - &
  - &
  - &
  -
  \\
$G_3$ &
  - &
  0.25 (W) &
  0.25 (W) &
  0.82 (S) &
  - &
  - &
  - &
  - &
  - &
  - &
  - &
  -
  \\
$G_4$ &
  - &
  0.25 (W) &
  0.25 (W) &
  0.82 (S) &
  - &
  - &
  - &
  - &
  - &
  - &
  - &
  -
  \\
$G_5$ &
  - &
  - &
  - &
  - &
  0.59 (N) &
  0.36 (W) &
  0.42 (N) &
  0.55 (N) &
  - &
  - &
  - &
  -
  \\
$G_6$ &
  - &
  - &
  - &
  - &
  0.55 (N) &
  0.24 (O) &
  0.35 (W) &
  0.55 (N) &
  - &
  - &
  - &
  -
  \\
$G_7$ &
  - &
  - &
  - &
  - &
  0.64 (N) &
  0.6 (N)  &
  0.55 (N) &
  - &
  - &
  - &
  - &
  -
  \\
$G_8$ &
  - &
  - &
  - &
  - &
  - &
  - &
  - &
  0.55 (N) &
  - &
  - &
  - &
  -
  \\
$G_9$ &
  - &
  - &
  - &
  - &
  0.59 (N) &
  0.36 (W) &
  0.42 (N) &
  0.59 (N) &
  - &
  - &
  - &
  -
  \\
$G_{10}$ &
  - &
  - &
  - &
  - &
  - &
  - &
  - &
  - &
  0.35 (W) &
  0.55 (N) &
  - &
  -
  \\
$G_{11}$ &
  - &
  - &
  - &
  - &
  - &
  - &
  - &
  - &
  0.35 (W) &
  - &
  - &
  -
  \\
$G_{12}$ &
  - &
  - &
  - &
  - &
  - &
  - &
  - &
  - &
  - &
  - &
  0.25 (W) &
  -
  \\
$G_{13}$ &
  - &
  - &
  - &
  - &
  - &
  - &
  - &
  - &
  - &
  - &
  - &
  0.13 (O)
  \bigstrut\\
\Xhline{4\arrayrulewidth}
\end{tabular}}%

%% file: table_relaxed.tex
\rowcolors{1}{}{lightgray}
\resizebox {0.9\textwidth }{!}{\begin{tabular}{ll}
\Xhline{4\arrayrulewidth}
Requirement &
  Sample RELAX-ed Value
  \bigstrut\\
\Xhline{4\arrayrulewidth}
$R_1$: achieve request transaction code &
  [expiry rate] as close as possible to $0.82 \times OV_1$
  \bigstrut\\
$R_2$: achieve latency examination &
  [examination delay] as close as possible to  $0.25 \times OV_2$
  \\
$R_3$: achieve one-time pad &
  [randomness] as close as possible to $0.25 \times  OV_3$
  \\
$R_4$: achieve SSL &
  [entropy] as close as possible to $0.82 \times  OV_4$
  \\
$R_5$: achieve password trial limitation &
  [trial delay] as close as possible to $0.59 \times OV_5$
  \\
$R_6$: achieve password policy &
  [complexity] as close as possible to $0.36 \times  OV_6$
  \\
$R_7$: achieve password encryption &
  [length of encryption key]as many bits as $0.42 \times  OV_7$
  \\
$R_8$: achieve random id &
  [randomness] as close as possible to $0.55 \times  OV_8$
  \\
$R_9$: achieve CAPTCHA &
  [level of distortion] as close as possible to $OV_9$
  \\
$R_{10}$: achieve complex pin &
  [complexity] as close as possible to $0.55 \times  OV_{10}$
  \\
$R_{11}$: achieve access control &
  [complexity] as close as possible to $0.25 \times  OV_{11}$
  \\
$R_{12}$: achieve redundant server &
  [number of servers] as close as possible to $0.13 \times  OV_{12}$
  \bigstrut\\
\Xhline{4\arrayrulewidth}
\end{tabular}}%

%% file: conclusion.tex
\section{Conclusions and Future Work}
\label{conclusion}

We presented a scalable framework referred to as the PAPS for goal-based prioritization and selection selection of security requirements which enhances precision of prioritization and selection by considering partiality of security. The proposed framework takes the security model of a software as the input and infers linguistic priorities of security requirements using a fuzzy inference system. Satisfaction conditions of security requirements then will be RELAX-ed when tolerated to partially selected (satisfy) those requirements. The defuzzified priority of a security requirement specifies its relaxed satisfaction level. Partial selection (satisfaction) of security requirements helps reduce the number of ignored security requirements which will ultimately reduce the number of unattended security threats. Validity of the PAPS framework was verified through studying an online banking system which served as our running example.  

The work is being continued by developing a tool based on the PAPS framework to assist automated prioritization and partial selection of security requirements. Using that tool in real world software project helps evaluate the PAPAS framework and its impact on the overall security of software in a real settings.  